\documentclass[10pt]{article}
\usepackage{amsmath, amssymb}
\usepackage{amscd}
\usepackage[dvips]{color,graphicx}
\usepackage{eucal}

\newtheorem{theor}{Theorem}[section]
\newtheorem{lem}{Lemma}[section]
\newtheorem{propo}{Proposition}[section]

\newenvironment{pf}{{\it Proof.}}{\hfill $\square$\\}

\title{\bf   Higher order approximation of isochrons}
\author{Daisuke Takeshita\footnote{University of Missouri-St. Louis, Department of Physics and Astronomy, Center for Neurodynamics, One University Blvd., St. Louis, MO 63121}, Renato Feres \footnote{Washington University, Department of Mathematics, Campus Box 1146, St. Louis, MO 63130}}

\begin{document}
\maketitle

\begin{abstract}
Phase reduction is a commonly used techinque for analyzing stable oscillators, particularly
in studies concerning synchronization and phase lock of
a network of oscillators. In a widely used numerical approach for
obtaining phase reduction of a single oscillator, 
one needs to obtain the gradient of the phase function, 
which essentially provides a linear approximation of isochrons. 
In this paper, we extend the method for obtaining partial derivatives 
of the phase function to arbitrary order, providing higher order 
approximations of isochrons. In particular, our method in order $2$ 
can be applied to the study of dynamics of a stable oscillator 
subjected to stochastic perturbations, a topic that will be 
discussed in a future paper.
We use the Stuart-Landau oscillator 
 to illustrate the method in order $2$.
 \end{abstract}

\section{Introduction and statement of main results}
Weak perturbations of limit cycle oscillators are of great interest
in a variety of fields in physics, chemistry, engineering, and quantitative biology,
whenever  the system under study displays stable oscillations.
A powerful theoretical approach in the analysis of weakly perturbed limit cycles,
particularly in relation  to synchronization and phase lock of a network
of oscillators,
is to reduce the description of the system to  a single ``phase'' variable.
This phase reduction procedure is the focus of the present paper.

In this introduction, we begin by describing a standard   numerical
approach for obtaining first order phase reduction, due to I.G. Malkin 
\cite{mal1,mal2}. 
We then explain our   higher
order method, which is summarized in Theorem \ref{main_theorem},  and illustrate it  in    order $2$
 using the well-known Stuart-Landau oscillator as  an example.  
 Proofs are
 are given in the subsequent sections. Our focus is on the theoretical underpinnings of the method 
 rather than on the details of numerical implementation, but 
   we provide a numerical example to illustrate the approach.
 
 The following set-up is assumed to hold throughout. 
  Let $F$ be a smooth (i.e., continuously differentiable to all orders) vector field on
 $n$-dimensional Euclidean space $\mathbb{R}^n$. The flow line of
 an initial point $x$ will be denoted 
 $\phi_t(x)$, or simply $x(t)$,  for $t\in \mathbb{R}$.  
 Let 
 $\mathcal{C}$  be a    stable (hyperbolic) limit cycle of the differential equation
 $\dot{x}=F(x)$ having period $T>0$. 
 We   write  $\omega=1/T$ for the reciprocal of the period. 
 The stability condition means that 
 for any given point $x$ on $\mathcal{C}$, there exists a 
 linear $n-1$-dimensional subspace $W(x)$ in  $\mathbb{R}^n$ transverse to $F(x)$  
 such that vectors in $W(x)$ contract exponentially under positive  iterations of
 the differential $(d\phi_T)_x$ of the flow map at $x$. That is, 
 $ |(d\phi_T)^n v| <C \lambda ^n |v|$
 for positive constants $C$,  $\lambda<1$, positive integer $n$, and    
  all $v$ in $W(x)$.
 As  we are only  concerned with the system near
$\mathcal{C}$, there is no loss of generality in assuming that $F$ is a complete vector
field, so that the flow lines $\phi_t(x)$ are defined for all
$t\in \mathbb{R}$ and all  $x\in \mathbb{R}^n$.
  
\subsection{The phase function}  
 We briefly review a few well-known  facts about  the   dynamics
  near a stable limit cycle for the purpose of setting up   notation. 
 
  Due to normal hyperbolicity on $\mathcal{C}$, it is known
  \cite{hps}
  that  a neighborhood of the limit cycle  is continuously  foliated by
  contracting manifolds, $\mathcal{W}(x)$, for each $x\in \mathcal{C}$,
  where the  $\mathcal{W}(x)$  are  smooth submanifolds  tangent to
  $W(x)$ at $x$ and diffeomorphic to
  an open disc. The foliation is invariant, i.e., $\phi_t$ maps $\mathcal{W}(x)$
  into $\mathcal{W}(\phi_t(x))$ for all $t\geq 0$ and all $x$ in $\mathcal{C}$. 
  Furthermore, the foliation is smooth since 
    $\mathcal{C}$ is an
  orbit of a smooth  flow.  
  In fact, we can use maps of the form $H(z, t)=\varphi_t(z)$,
  where $z\in \mathcal{W}(x)$ and $t\in (-a, a)$,
  to produce smooth foliation charts around $x$.

  Let  $\Theta$ be a  function
  defined on a sufficiently small neighborhood of $\mathcal{C}$,
  whose level sets are the local contracting manifolds and such that 
  $\Theta(x(t))=\omega t$ modulo integer translations. I.e., $\Theta$ takes
  values in $\mathbb{R}/\mathbb{Z}$. We   often regard
  $\Theta$ simply as a   function into $\mathbb{R}$,
  keeping in mind that, in this case, $\Theta(x(t+nT))=\Theta(x(t))+n$. Derivatives of $\Theta$,
  on the other hand, are single-valued functions into $\mathbb{R}$.

We refer to $\Theta$ as the {\em phase function} of the oscillator. It is, thus,
    a smooth function on some neighborhood of the limit cycle such that
$\Theta(y)=\Theta(x)$ iff  $\vert \phi_t(y)-\phi_t(x)\vert \to 0$ 
as $t \to \infty$ whenever
   $x\in \mathcal{C}$ 
and y  lies  in a sufficiently small  neighborhood of $\mathcal{C}$.

 The level sets of $\Theta$ are also called {\em isochrons} \cite{winfree}.
The existence of isochrons was first proved  in  \cite{guck}. 
Since the  component functions    of the gradient  of $\Theta$
 give
the change in  phase due to a small change  in the respective  position 
coordinates \cite{br}, 
the graphs of those functions are often referred to as
 {\em phase response curves}. This notion is widely used in
  theoretical and experimental neuroscience. 
 For example, some theoretical studies have examined how the shape of the phase response 
 curves affect the sychronization dynamics of 
 coupled oscillators \cite{hansel, ermenPRC}. 
 Other studies have investigated the connection between 
  phase response curves and mechanisms of brain function 
 and pathology such as autoassociative memory \cite{lengyel} 
 and epileptic seizures \cite{take}. 
The phase response curves of neurons in the neocortex have been experimentally determined \cite{netoff, tsubo}.

\subsection{First order phase reduction}
Before explaining  our main results, which
are collected  in Theorem \ref{main_theorem}, it is useful to briefly review the standard method
of phase reduction. For a more complete discussion the reader is referred to 
 \cite{iz}, Chapter 10,  and to \cite{br}.

It is easy to see why the derivatives of $\Theta$ are needed
when studying weak perturbations of a stable oscillator described by 
$\dot{x}=F(x)$. 
Let such a perturbation be given  by the new equation 
$\dot{x}=F(x)+\epsilon G(x,t)$, where $\epsilon$ is a small positive number. Then
\begin{equation}
\frac{d}{dt}\Theta(x(t))={\nabla}\Theta(x(t))\cdot \dot{x}
=\omega + \epsilon {\nabla}\Theta(x(t))\cdot G(x,t)
\end{equation}
where ${\nabla}\Theta $ is the gradient of the phase function.
Writing $Q^{(1)}_t={\nabla}\Theta(x_0(t))$,  where $x_0(t)$ is
the point on the limit cycle on the same isochron as $x(t)$,  then
\begin{equation}
\frac{d \Theta}{dt}=\omega + \epsilon Q^{(1)}_t \cdot G(x_0(t),t) + \epsilon \mathcal{O}(|x(t)-x_0(t)|). \label{1stordeq}
\end{equation}
One then proceeds by discarding 
 the  term $\epsilon \mathcal{O}(|x(t)-x_0(t)|)$
 and analyzing the resulting system.  Thus
 implementing the phase reduction method  in order $1$
 in $|x(t)-x_0(t)|$
  requires  finding
  the gradient of the phase function along
 the limit cycle of the unperturbed oscillator,  denoted 
 by $Q^{(1)}_t$ in the last equation.

One practical method for obtaining  the gradient of $\Theta$ is by solving the   equation:
\begin{equation}
\dot{Q}^{(1)}_t+DF^{\dagger}(x_0(t))Q^{(1)}_t=0
\end{equation}
where the dot indicates time derivative and 
$DF^{\dagger}(x_0(t))$ is the transpose  of 
the Jacobian matrix of $F$ evaluated 
along  the limit cycle.  
 This  procedure was suggested by Malkin \cite{hop,mal1, mal2} and later by others independently \cite{neu, ermkop}. The reader is referred to \cite{hop}, Chapter 9,  for more details of the Malkin's theorem and \cite{iz}, Chapter 10 for a historical note on phase reduction.
One can find $Q^{(1)}_t$  
by numerically integrating the equation backwards in time for any initial condition
satisfying $Q_0^{(1)}\cdot F(x_0(0))=1$, over an interval of time long enough
to allow the solution to stabilize to a periodic one \cite{intback}.

One limitation of the method as presented above is that it
 is only valid to first order. To develop higher order phase reduction, 
 it is necessary to obtain higher order partial derivatives of 
 $\Theta$. However, to the best of our knowledge, 
an approach to higher order approximations
of $\Theta$ similar in spirit to the above due to Malkin  for finding
the gradient of $\Theta$ has not been described so far 
in the literature. The goal of the present paper is to develop 
a numerical method to obtain partial derivatives of $\Theta$ to arbitrary order.
 
 Even within the framework of first order phase reduction 
 (\textit{i.e.} Equation \eqref{1stordeq}), 
 in certain situations, as when dealing with weak stochastic perturbations of
 oscillators, one may need to know the  derivatives of $\Theta$
 along the limit cycle to orders  greater than $1$. 
 For example, in \cite{prl},   a stochastic version of 
 phase reduction is given using the 2nd order partial derivatives 
 of $\Theta$ to obtain the mean and variance of the period 
 of a limit cycle oscillator. 
 They   apply their results to 
   the Stuart-Landau oscillator, for which an explicit form of $\Theta$ 
 can be obtained analytically. In general,  however, an analytical form of $\Theta$ 
 is not available.
 Therefore, the method we present here  is of particular interest in
 order two  for studying the   dynamics of a limit cycle 
 oscillator subjected to stochastic perturbations. This 
  will be discussed in detail in a future paper.

The main result of this paper, which
is described in Theorem \ref{main_theorem} below,
 amounts to  a recursive procedure for finding
 the higher order derivatives of $\Theta$ in which the first step is  Malkin's method just described.
 
 \subsection{The main result} \label{main_result}
 We need a few definitions first. 
 Let $v=(v_1, \dots, v_n)$ be any vector in $ \mathbb{R}^n$ and 
 $f$  any differentiable real-valued function in   
 $x_1, \dots, x_n$. 
 Given a multi-index $J=(j_1, \dots, j_n)$, i.e., an $n$-dimensional vector
 with non-negative integer entries,
 we write $v^J=v_{j_1} \dots  v_{j_n}$ 
 and
  $f_J=D^Jf=D_1^{j_1}\dots D_n^{j_n}f$, 
  where $D_i$ denotes partial derivative with
 respect to $x_i$ and $D_i^k=D_i \dots D_i$ ($k$ times).
 Let  $f^{(k)}$ represent the symmetric $k$-multilinear  map on $\mathbb{R}^n$ 
 characterized (via polarization of polynomial maps) by
$$f^{(k)}(v, \dots, v)=\sum_{|J|=k}f_{J}v^J$$
for all $v\in \mathbb{R}^n$. The sum
 is over all multi-indices $J$ of  order 
$|J|=j_1 + \dots +j_n=k$. Here, and often later, we
  omit reference to the point  $x$  where the derivatives are taken. When
  necessary,    this point is   indicated  as
  a sub-index; thus $f_x^{(2)}(v,w)$ is the bilinear map evaluated
  at the   vectors $v, w$ regarded as tangent vectors at $x\in \mathbb{R}^n$,
  where $x$ is the point where the partial derivatives of $f$ are calculated. 

 We now  define 
  \begin{equation}\label{defi}Q_t^{(k)}=\Theta^{(k)}_{\phi_t(x)},\end{equation}
  for some fixed  $x\in \mathbb{R}^n$,
where $\phi_t$ is the flow of $F$.
Similarly, we define $F^{(k)}$, which is now a vector valued, symmetric,
$k$-multilinear  map.
(A convenient alternative description of $\Theta^{(k)}$ and $F^{(k)}$,
and more generally of the higher order derivative forms
associated to  tensor fields,  will be
given later in the paper.) In particular, $F^{(1)}$ is the linear map which  
to   $v\in \mathbb{R}^n$ gives  the directional derivative of $F$ along $v$, i.e.,
 $F^{(1)}(v)=D_vF=\sum_j v_jD_jF$, where $D_vF$ is defined by this identity.

Another general concept needed below is the {\em symmetric composition} of multilinear 
maps, which we define as follows.
Let $Q$ by a symmetric $s$-multilinear  map on $\mathbb{R}^n$ taking values in $\mathbb{R}$,
and $H$ a symmetric $k$-multilinear  map on $\mathbb{R}^n$ taking values in $\mathbb{R}^n$. 
Then the symmetric composition of $Q$ and $H$ is the symmetric $s+k-1$-multilinear
map on $\mathbb{R}^n$, denoted  $Q \odot H$ and given by
$$ Q\odot H(v_1, \dots, v_{s+k-1})=\frac1{(s+k-1)!}\sum_{\sigma\in S_{s+k-1}}Q(H(v_{\sigma(1)}, \dots, v_{\sigma(k)}), v_{\sigma(k+1)}, \dots, v_{\sigma(s+k-1)})$$
where the sum is over all permutations of the set $\{1, 2, \dots, s+k-1\}$.
Finally, given a co-vector $Q$ on $\mathbb{R}^n$ 
(a linear map from $\mathbb{R}^n$ to $\mathbb{R}$), 
we define the $k$-multilinear map
$$ (Q\otimes \cdots \otimes Q)(v_1, \dots, v_k)=Q(v_1) \dots Q(v_k)
\textrm{ (the }k\textrm{-fold tensor product.)}$$

\begin{theor}\label{main_theorem}
Let $x \in \mathcal{C}$, $x(t)=\phi_t(x)$, $k\geq 1$,  and 
$Q^{(k)}_t$  the $k$-multilinear map defined in \eqref{defi}. 
Then, the following hold.
\begin{enumerate}
\item   
$Q^{(k)}_t$ satisfies the  differential equation in $A_t^{(k)}$ given by
\begin{equation}\label{eq_theorem}
\dot{A}_t^{(k)} + k A_t^{(k)}\odot F^{(1)}_{x(t)}= - \sum_{l=1}^{k-1} \binom{k}{l+1}Q_t^{(k-l)}\odot F^{(l+1)}_{x(t)},
\end{equation}
where the right-hand side involves the $Q^{(j)}_t$ for $j<k$, and equals $0$ 
if $k=1$.
\item
Let A be any $k$-multilinear map, $N$ a positive integer, and
$A_{t,N}$ the solution to Equation \eqref{eq_theorem} 
for $0 \le t \le NT$ such that $A_{NT,N}=A$. Then there exists a T-periodic solution $A_t$ such that $A_{t,N}$ converges exponentially to $A_t$ for $0\leq t\leq T$  as $N \to \infty$.
More precisely, there are constants $C>0$ and $0<\lambda<1$ so that 
$$\sup_{0 \le t \le T}\vert A_{t,N}-A_{t} \vert \le C\lambda^N.$$
\item
If $A^{(k)}_t$ is any $T$-periodic solution of Equation \eqref{eq_theorem}, then
$Q^{(k)}_t=A^{(k)}_t+\mu Q^{(1)}_t \otimes \cdots \otimes Q^{(1)}_t $
where $$\mu=T^k[Q^{(k)}_{0}(F_x,\dotsc,F_x)-A^{(k)}_{0}(F_x,\dotsc,F_x)].$$
\item 
The term $Q^{(k)}_{0}(F_x,\dotsc,F_x)$ has an {\em a priori} expansion  as a linear combination of compositions
of  the lower order terms
$Q^{(l)}_{0}$ and $F^{(l)}_x$ for $l\leq k-1$. This expansion is described
in section \ref{forest}.
\end{enumerate}
\end{theor} 
 
 The equation for the standard   (Malkin's) method 
of phase reduction for obtaining the gradient of $\Theta$ is  the equation 
in part 1 of the theorem when $k=1$:
\begin{equation}\label{malkin1}
\dot{Q}_t^{(1)} + Q^{(1)}_t\circ F^{(1)}_{x(t)}=0.
\end{equation} 
Furthermore, 
$Q_t^{(1)}$ is the unique periodic solution such that $Q_0^{(1)}(F_x)=1/T$.
This unique solution can be found numerically, according to part 2,  by the following procedure:
Let a covector $A$ (i.e., a linear map from $\mathbb{R}^n$ to $\mathbb{R}$) at $p$ be a  choice of initial condition for 
Equation \ref{malkin1} which is arbitrary except for the normalization
$A(F_p)=1/T$.  One then integrates Equation \eqref{malkin1}  for $t<0$ (backward integration)
until the solution stabilizes to a periodic (co)-vector-valued function on
the limit cycle.  Stabilization is assured to happen for sufficiently large $|t|$.
This periodic function is  the solution we want in order one. 
The general order case is then given recursively  by the  successive applications of the  theorem.

 Before presenting the proof, we illustrate the use of the theorem with the
 Stuart-Landau oscillator. 

\subsection{The Stuart-Landau oscillator: an illustration}
To illustrate the method, we focus attention on the
case $k=2$. From the general definition of symmetric composition introduced above we have
that 
$Q^{(2)}_t \odot F^{(1)}$ and   $Q^{(1)}_t \odot F^{(2)}$  are given by
 \begin{align*}
 Q^{(2)}_t \odot F^{(1)}(v_1, v_2)&=\frac12\left(Q^{(2)}_t(F^{(1)}(v_1), v_2) +  Q^{(2)}_t(F^{(1)}(v_2), v_1)\right)\\
 Q^{(1)}_t \odot F^{(2)}(v_1, v_2)&=Q^{(1)}_t (F^{(2)}(v_1, v_2)).
 \end{align*}

We suppose that $Q^{(1)}_t$ has already been obtained  (say, by the standard method)
 and wish to find
$Q^{(2)}_t$. According to the main theorem, this second order term satisfies the non-homogeneous differential equation
\begin{equation}\label{malkin2}
\dot{Q}^{(2)}_t +2 Q^{(2)}_t\odot F^{(1)}=- Q^{(1)}_t  \odot F^{(2)},
\end{equation}
where the $F^{(j)}$ are evaluated at $x(t)$ on the limit cycle. 
The equation can be solved  as follows:
 Let $A^{(2)}_t$ be a   solution to Equation \ref{malkin2}
 obtained by backward integration for an  arbitrary initial condition.  For large enough
 $|t|$ this solution stabilizes to a periodic (tensor-valued) function along the
 limit cycle, which we still denote by $A^{(2)}_t$. Then, by   item 3 of Theorem \ref{main_theorem},
 \begin{equation}\label{solution2_0}
 Q_t^{(2)}=A^{(2)}_t + T^2\left(Q_0^{(2)}(F_x,F_x)-A_0^{(2)}(F_x,F_x) \right)Q_t^{(1)}\otimes Q^{(1)}_t
 \end{equation}
where
 $Q_t^{(1)}\otimes Q^{(1)}(v_1, v_2)=Q^{(1)}_t(v_1) Q^{(1)}_t(v_2).$ 
We   show later that the expansion referred to in item 4 of the theorem 
amounts in this case to 
$Q_0^{(2)}(F_x,F_x)=-Q_0^{(1)}(F^{(1)}_x(F_x))$. Therefore,
\begin{equation}\label{solution2}
Q_t^{(2)}=A^{(2)}_t - T^2\left(Q_0^{(1)}(F^{(1)}(F_x))+A_0^{(2)}(F_x,F_x) \right)Q_t^{(1)}\otimes Q^{(1)}_t
 \end{equation}
is the solution we seek.

We now  recall the Stuart-Landau oscillator. (See \cite{prl}.) Define
$$A_a=\left[\begin{array}{rr}1 & -a \\ a & 1\end{array}\right].$$
 We regard points of $\mathbb{R}^2$ as column vectors: $x=(x_1, x_2)^\dagger$, where $\dagger$ indicates transpose.
Let  $a, b$ be real constants and $\rho(r)$    a smooth function of $r>0$ such that $\rho(1)=1$
and $\rho'(1)=\chi>0$. 
Now define a vector field on $\mathbb{R}^2$ by
\begin{equation}\label{sl}
F(x)=A_ax - \rho(|x|) A_bx.
\end{equation}
Then it is easy to check that the differential equation $\dot{x}=F(x)$ has a hyperbolic
stable limit cycle given by $S^1=\{x\in \mathbb{R}^2: |x|=1\}$. In fact, $r=|x|$
satisfies:
$$\dot{r}=r(1-\rho(r))=-\chi (r-1) +o(r-1),$$
showing that the limit cycle is approached for $t>0$ with Lyapunov exponent $-\chi$.
Specializing to  $\rho(r)=r^2$, then $\dot{r}=r(1-r^2)$ is easily solved:
\begin{equation}\label{solr}
r(t)=\left(1+\frac{1-r_0^2}{r_0^2}e^{-2t}\right)^{-\frac12},
\end{equation}
where $r_0=r(0)$.
With the coordinate change
 $x_1=\cos(\varphi + b \ln r)$ and $ x_2 =\sin(\varphi + b \ln r)$ we can write the solution
 to $\dot{x}=F(x)$
 explicitly in the new variables $r, \varphi$ by setting   
 \begin{equation}
 \varphi(t)=\varphi_0 +(a-b)t,
 \end{equation}
 as can be easily checked.
 Therefore, $\Theta(x)=\varphi /2\pi$ modulo integer translations.
For this example, we can calculate the derivatives of $\Theta(x)$
explicitly, and then compare them with the numerical values  derived from Theorem \ref{main_theorem}.

Implicit differentiation   gives the first and second order derivatives of $\Theta$
along the limit cycle. We write   $\Theta_i=D_i\Theta$, $\Theta_{ij}=D_iD_j\Theta$, where
$D_i$ is partial derivative in $x_i$. Then  
\begin{equation}
 \Theta^{(1)} =  (\Theta_1, \Theta_2)=\left(-\frac{bx_1+x_2}{2\pi|x|^2},\frac{x_1-bx_2}{2\pi|x|^2} \right).
\end{equation} 
 Identifying $\Theta^{(2)}$ with the Hessian  of $\Theta$, we can write
 \begin{equation}
 \renewcommand{\arraystretch}{1.6}
  \Theta^{(2)} = 
  \left[\begin{matrix}\Theta_{11} & \Theta_{12} \\\Theta_{21} & \Theta_{22}\end{matrix}\right]=
  \left[\begin{matrix} \frac{2x_1x_2 - b(x^2_2-x^2_1)}{2\pi|x|^4} &\ \ \frac{2bx_1x_2 + x_2^2 - x_1^2}{2\pi|x|^4} \\ \frac{2bx_1x_2 + x_2^2 - x_1^2}{2\pi|x|^4} & 
  -\frac{2x_1x_2 - b(x^2_2-x^2_1)}{2\pi|x|^4}\end{matrix}\right].
 \end{equation}
The tensors $Q^{(1)}_t$ and $Q^{(2)}_t$ are similarly written. Let 
$\zeta(t)= \varphi_0 + (a-b)t + b\ln r(t)$. Then
\begin{equation}
  Q^{(1)}_t =  (Q_1(t),Q_2(t))= \left(\frac{\sin\zeta(t)}{2\pi r(t)}, \frac{\cos\zeta(t)}{2\pi r(t)}\right)
\end{equation}
and 
\begin{equation}
\renewcommand{\arraystretch}{1.4}
 Q^{(2)}_t = \left[\begin{matrix}Q_{11}(t) &Q_{12}(t) \\ Q_{21}(t)  &Q_{22}(t) \end{matrix}\right]=\left[\begin{matrix}
\frac{b\cos(2\zeta(t)) + \sin(2\zeta(t))} {2\pi r^2(t)} & \frac{-\cos(2\zeta(t)) + b\sin(2\zeta(t))} {2\pi r^2(t)} \\ 
\frac{-\cos(2\zeta(t)) + b\sin(2\zeta(t))} {2\pi r^2(t)} & -\frac{b\cos(2\zeta(t)) +\sin(2\zeta(t))} {2\pi r^2(t)}\end{matrix}\right].
\end{equation}
For any vectors $v, v_1, v_2\in \mathbb{R}^2$,  
\begin{align}\label{FFF1}
F&=A_ax-|x|^2A_bx\\
\label{FFF2}
F^{(1)}(v)&= A_a v -2x\cdot v  A_b x - |x|^2 A_bv\\
\label{FFF3}
F^{(2)}(v_1, v_2)&=   -2x\cdot v_1 A_b v_2 - 2 x\cdot v_2 A_b v_1  -2 v_1\cdot v_2 A_b x.
\end{align}
Let the components of these tensors relative to the standard basis $e_1, e_2$ of $\mathbb{R}^2$  be  denoted as follows:
\begin{equation*}
\renewcommand{\arraystretch}{1.4}
F_x=\left[\begin{matrix}F^1(x) \\ F^2(x) \end{matrix}\right], \ \ 
 F^{(1)}_x(e_i)=\left[\begin{matrix}F^1_i(x) \\ F_i^2(x) \end{matrix}\right], \ \  F_x(e_i,e_j)=\left[\begin{matrix}F_{ij}^1(x) \\ F_{ij}^2(x) \end{matrix}\right],
\end{equation*}
where the entries are obtained from Equations \ref{FFF1}, \ref{FFF2}, and \ref{FFF3}. For example,
from Equation \ref{FFF3} it follows that  $F^2_{21}(x)=F^2_{12}(x)=e_2\cdot F^{(2)}(e_1, e_2)=-2(x_1 +bx_2).$ The other entries are:

\begin{align*}
\renewcommand{\arraystretch}{1.4}
\left[\begin{matrix}F_1^1(x) & F^1_2(x) \\ F_1^2(x)& F_2^2(x)  \end{matrix}\right]&=
\renewcommand{\arraystretch}{1.4}
\left[\begin{matrix}1- 2x_1(x_1-bx_2)- |x|^2& -b-2x_2(x_1-bx_2) + b|x|^2 \\ 
b-2x_1(bx_1+x_2) - b|x|^2& 1- 2x_2(bx_1+x_2)- |x|^2 \end{matrix}\right]\\
\renewcommand{\arraystretch}{1.4}
\left[\begin{matrix}F_{11}^1(x) & F^1_{12}(x) \\ F_{21}^1(x)& F_{22}^1(x)  \end{matrix}\right]&=
\renewcommand{\arraystretch}{1.4}
\left[\begin{matrix}-6x_1 + 2bx_2&  2b x_1 -2x_2\\ 
2b x_1 -2x_2 & -2x_1 + 6 b x_2\end{matrix}\right]\\
\renewcommand{\arraystretch}{1.4}
\left[\begin{matrix}F_{11}^2(x) & F^2_{12}(x) \\ F_{21}^2(x)& F_{22}^2(x)  \end{matrix}\right]&=
\renewcommand{\arraystretch}{1.4}
\left[\begin{matrix}-6bx_1 - 2x_2&  -2 x_1 -2bx_2\\ 
-2 x_1 -2bx_2 & -2bx_1 - 6  x_2\end{matrix}\right]
\end{align*}

\begin{figure}[htbp]
\begin{center}
\includegraphics[width=4.5in]{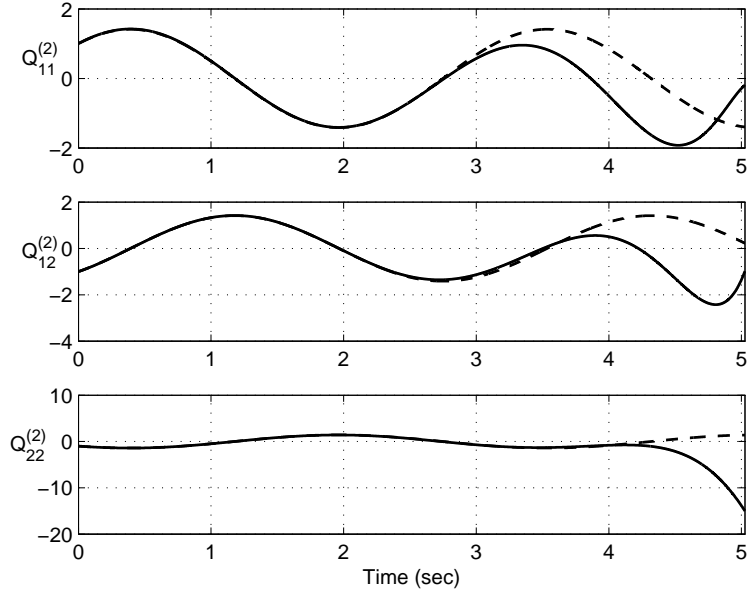}
\caption{\small Determination of $Q^{(2)}_t$ 
for the Stuart-Landau oscillator with $a=2, b=1$. 
Each component of the tensors is plotted as a function of time. 
Numerical integration is done backward in time using the Euler method with $dt=10^{-3}$.
Solid line: $Q^{(2)}_t$ 
determined from $A_{t,NT}$;
dashed line: $Q^{(2)}_t$ found analytically.
To clearly show   convergence of the numerically obtained $Q^{(2)}_t$ to 
the analytical solution, $A_{t,NT}$ was not plotted and the period of the simulation was set at only $0.8T$.}
\label{fig:SL}
\label{tensor} 
\end{center}
\end{figure}

We can now write the 
differential equations for $Q_i$ and $Q_{ij}$ (Equations \ref{malkin1} and \ref{malkin2}; 
all summations are over $s=1,2$):

\begin{align}
\dot{Q}_i&= \sum_s Q_sF^s_i\\
\dot{Q}_{ij}&=\sum_s(F_i^sQ_{sj}+ F^s_jQ_{si})  - \sum_s Q_s F^s_{ij}
\label{eq_Q2}.
\end{align}

The first equation is the one used in the standard Malkin's approach. Once it is
solved by the already indicated procedure, its solution enters as the
non-homogeneous term for the second equation. 

The result of the numerical calculation is  shown in Figure \ref{fig:SL}. 
The initial condition for $A_{t,NT}$ was chosen from a set of random numbers 
and numerical integration of Equations \eqref{eq_Q2} 
was done backward in time (data not shown). 
Then, $Q^{(2)}_t$ was obtained using Equation \eqref{solution2_0} 
(solid line in Figure \ref{fig:SL}). Convergence of $Q^{(2)}_t$, 
obtained numerically, to the analytical solution
(dashed line in Figure \ref{fig:SL}) is  clearly observed.

\section{Proof of the main theorem}
The subsequent sections are dedicated to proving Theorem \ref{main_theorem}.
It is convenient, for reasons  we hope will become apparent 
along the way,   to use a more  geometric and coordinate-free
language for manipulating tensors, even though all calculations are done in $\mathbb{R}^n$.
This  preparatory material on tensor calculus  will take a few  pages to develop, so it
may be appropriate to provide an overview of the proof in simpler terms first. 

The basic idea  for proving    part 1 of Theorem \ref{main_theorem} 
is the following.
  An immediate consequence of
  the definition of the phase function is that $\Theta(\phi_t(\gamma(s)))=\omega t + \Theta(\gamma(s))$ for any differentiable curve $\gamma(s)$. Therefore,
$\frac{d}{dt} \frac{d^k}{ds^k}\Theta(\phi_t(\gamma(s)))=0$ for $k\geq 1$. This  gives an ODE that the $k$th derivative of the phase function must  satisfy. Thus the proof of part 1
simply amounts  to successive applications of elementary differentiation rules, although
handling the multitude of terms that arise quickly becomes a challenge.
 In fact, this is precisely the kind of situation that
often calls for the use of Fa\`a di Bruno type formulas. We take here a different approach,
which obtains   the differential equation recursively and uses a coordinate-free language
 to facilitate the manipulation of terms.
 
 The proof  also involves   a somewhat  surprising cancellation of terms that
  renders the result more simple  than  might be  expected at first. 
To better understand this point, we sketch the derivation of the differential
equation  
in dimension one. 
Although the calculation is considerably
more complicated in the general case,
the key formal aspects of the derivation   are already present for $n=1$. 
We use subscripts to indicate the number of derivatives, so $f_k(x)$ is
the $k$th derivative of   $f(x)$, whereas $f'(x)$ is
also used for the first derivative. Let $\gamma(s)=x_0 + vs$, where $x_0$
and $v$ are fixed.  Let
$u_t(s)=\phi_t(\gamma(s))$, so $u_t'(0)=\phi_t'(x_0)v$.
Also set $Q^{(k)}_t(s)=\Theta_k(u_t(s))$. Recall that we wish to find
a differential equation for $Q^{(k)}_t=Q^{(k)}_t(0)$. The differential equation
associated to   $F$   is $\dot{u}_t(s)=F(u_t(s))$.
Then for $k=1$,  
\begin{equation*}
0=\frac{d}{dt} \frac{d}{ds}\Theta(\phi_t(\gamma(s)))
=\left\{ \dot{Q}^{(1)}_t(s)
+Q^{(1)}_t(s)F_1(u_t(s))\right\}\phi_t'(\gamma(s))v
\end{equation*}
for arbitrary $v$.
An exchange of the order of derivatives in $t$ an $s$ was used in the
second term of the right-hand side of the equation. Thus we conclude
that $\dot{Q}^{(1)}_t(s) 
+ F_1(u_t(s)) Q^{(1)}_t(s)=0$, from which the claim  for $k=1$ follows
by taking $s=0$.

We now suppose that $$\dot{Q}_t^{(j)}(s)+\sum_{l=1}^{j}\binom{j}{l}F_{l}(u_t(s))Q_t^{(j+1-l)}(s)=0 $$
for $1\leq j\leq k$ and wish to show that  this also  holds for $j=k+1$. 
Taking one more   derivative in $s$ and exchanging derivatives in $t$ and $s$ gives
\begin{equation*}
\begin{split}
0&=\frac{d}{ds}\left\{
\frac{d}{dt} \Theta_k(u_t(s)) +  \sum_{l=1}^k \binom{k}{l}F_l(u_t(s))\Theta_{k+1-l}(u_t(s))
\right\}\\
&= \left\{\dot{Q}_t^{(k+1)}(s) +  Q_t^{(k+1)}(s)F_1(u_t(s))\right\}u'_t(s)\\
&\qquad + \sum_{l=1}^k \binom{k}{l}\left\{F_{l+1}(u_t(s))Q_t^{(k+1-l)}(s)+
F_l(u_t(s))Q_t^{(k+2-l)}(s)\right\} u'_t(s).
\end{split}
\end{equation*}
Thus the desired   equation will hold for $j=k+1$ if we can show that
\begin{equation*}
\begin{split}
   Q_t^{(k+1)}(s)F_1(u_t(s)) 
 &+ \sum_{l=1}^k \binom{k}{l}\left\{F_{l+1}(u_t(s))Q_t^{(k+1-l)}(s)+
F_l(u_t(s))Q_t^{(k+2-l)}(s)\right\}\\
 &\qquad = \sum_{l=1}^{k+1}\binom{k+1}{l}F_l(u_t(s))Q^{(k+2-l)}_t(s). 
 \end{split}
 \end{equation*}
 This indeed holds 
 and is  easily checked for $n=1$ by using the 
 well-known relation 
  $$\binom{k}{l-1}+\binom{k}{l}=\binom{k+1}{l}.$$
 The    $n\geq 2$ case is considerably more
 involved but  follows a similar line of proof. 

Our use of tensor calculus becomes more essential in the proof of parts
2  and 3 of the main theorem, which rely  on  a study of the stability properties
of the tensor equation  given in part 1. Part 4 boils down to an enumeration of
 certain combinations of tensors that
can be represented  by 
formal linear combinations of rooted trees. This is discussed
later under the heading ``forest expansion.''

\subsection{Higher  order differentials of   tensor fields}
In this and the next two sections we develop some background 
material on  differential calculus
of tensor fields that will make our  calculations
of Taylor expansions of functions and vector fields   more tractable.
We suppose that  the  standard concepts (see, for example, \cite{kob}), such
as tangent and cotangent bundles,  Lie derivatives, covariant differentiation, etc,
are known, but  recall some of the definitions 
 for the purpose of setting up   notation. 

\begin{figure}[htbp]
\begin{center}
\includegraphics[width=.6in]{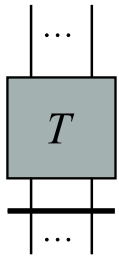}
\caption{\small  Diagram for the symmetrization of a $(r,k)$-tensor. The symmetrization
of the contravariant legs is indicated by a thick  bar. }
\label{tensor2} 
\end{center}
\end{figure}

 Let $V$ be a finite dimensional
vector space and $V^*$ its dual space. 
The space of tensors of type $(r,s)$ is $$V^{(r,s)}=V\otimes \cdots \otimes V\otimes V^*\otimes \cdots
\otimes  V^*,$$  in which  there are $r$ copies of $V$ and $s$ copies of $V^*$.
We   think of  an element of  $V^{(r,s)}$ as  an $s$-multilinear function taking values
in the set of $(r,0)$-tensors.  We find it useful to
represent a tensor $T$ of
type  $(r,s)$  as a diagram consisting of a box with $s$ (contravariant) lower legs   and $r$ (covariant) upper legs, and think of the lower legs as places where
the vector arguments  are plugged in.  See for example, Figure \ref{tensor2}.

A {\em tensor field} of type $(r,s)$ on $\mathbb{R}^n$
is a function that associates to each $x$ in $\mathbb{R}^n$ an element of
$V_x^{(r,s)}$, where $V_x=T_x\mathbb{R}^n$ is  the tangent space at $x$.
(This tangent space is, of course,  naturally identified with $\mathbb{R}^n$, 
but   observing the distinction will help keep track of
where derivatives are evaluated.) 

All the tensor fields (functions, vector fields,
etc.) that are considered below are smooth. The result of evaluating an $(r,s)$ tensor $T$ on
$s$ vectors $X_1, \dots, X_s$ is the $(r,0)$ tensor denoted $T(v_1, \dots, v_s)$.


Let $D$ 
be the standard covariant derivative in $\mathbb{R}^n$. If $X$ is
a vector field and $v$ is a tangent vector at some point, then ${D}_vX$ is
the standard derivative of $X$ along $v$. The derivative
of an  ordinary function $f$ along $v$ will be written as   $vf ={D}_vf $. We
can then extend the definition ${D}_v$ to general tensors by imposing the condition
that the product rule for differentiation holds for  the tensor product
$\otimes$  as well  as for  the pairing  operation of $V$ and $V^*$.  This implies that if $X_1, \dots, X_s$
are vector fields and $v\in T_x\mathbb{R}^n$, then
\begin{equation}\label{tensoreq}
({D}_v T)(X_1, \dots, X_s)={D}_v T(X_1, \dots, X_s)-\sum_{j=1}^sT(X_1, \dots, {D}_v X_j,\dots, X_s).\end{equation}
When $T$ has type $(0,s)$,  $T(X_1, \dots, X_s)$ is an ordinary function,
and we write $vT(X_1, \dots, X_s)$ instead of ${D}_v T(X_1, \dots, X_s)$.
The result of the differentiation of $T$ is an $(r, s+1)$ tensor field ${D} T$.

Similarly, ${D}^kT$ is the $(r, s+k)$ tensor
field obtained from $T$ by applying ${D}$ $k$-times. We call ${D}^kT$ the $k$-th {\em order differential} of $T$.
Regarding $T^{(k)}={D}^kT$ as a $k$-multilinear map taking values in 
the space of $(r,s)$ tensors, then it can also be defined inductively:
$T^{(1)}(v) = {D}_v T$ and
$T^{(k)}(v_1,\dots,  v_k)=({D}_{v_1}T^{(k-1)})(v_2, \dots, v_k)$ for $k\geq 2$.

We are dealing with the standard covariant differentiation in 
  $\mathbb{R}^n$,  so  ${D}$ is symmetric (i.e., its torsion tensor is 0) and flat (i.e., its curvature tensor is 0): 
 given vector fields $X_1, X_2, X_3$, then
${D}_{X_1}X_2-{D}_{X_2}X_1= [X_1, X_2]$
 and ${D}_{X_1}{D}_{X_2}X_3 -{D}_{X_2}{D}_{X_1}X_3={D}_{[X_1,X_2]}X_3$,
 where $[X_1, X_2]$ is the Lie bracket. 
 We also say that an $(r,s)$-tensor  field $S$ is {\em symmetric}
 if  for each  
  permutation $\sigma$ of the set $\{1, \dots, k\}$ and  each  $x$, 
$S_x^{(k)}(v_{\sigma(1)}, \dots, v_{\sigma(k)})=S^{(k)}_x(v_1, \dots, v_k)$ for
all vectors $v_1, \dots, v_k\in T_x\mathbb{R}^n$.  For simplicity, 
we are not indicating the $s$    vector arguments
of $S$.

\begin{propo}
Let $T$ be an $(r,s)$-tensor  field in $\mathbb{R}^n$ and $T^{(k)}$ the $k$-th order
differential of $T$. Then $T^{(k)}$ is symmetric.  
\end{propo}
\begin{pf}
We indicate the proof for $k=2$. The general case follows by   induction  
using the same argument. Let $X_1, X_2$ be
vector fields that agree with $v_1, v_2$ at $x$. If follows from the definitions that 
$$T^{(2)}_x(v_1, v_2)={D}_{X_1}{D}_{X_2}T- {D}_{{D}_{X_1}X_2}T$$
where the right-hand side is evaluated at $x$. 
Therefore, 
$$T^{(2)}_x(v_1, v_2)-T^{(2)}_x(v_2, v_1)=[{D}_{X_1},{D}_{X_2}]T- {D}_{[X_1,X_2]}T=0, $$
where the last equality is due to the fact that ${D}$ is flat.
\end{pf}

As an example, let $T$ be an ordinary function, denoted $f$.
Let $X_1, \dots, X_n$ be the coordinate vector fields in $\mathbb{R}^n$: $X_j=\frac{\partial}{\partial x_j}$.
 Then it is easy to check that
$$f^{(k)}(X_{j_1}, \dots, X_{j_k})=X_{j_1} \dots X_{j_k} f, $$
which is the $k$-th order partial derivative of $f$ with respect to $x_{j_1}, \dots, x_{j_k}$.

\subsection{Symmetric composition of tensors}
Let $S_k$ represent the group of permutations of the set $\{1, \dots, k\}$.
Given a tensor $T$ of type $(r,k)$, we define its {\em symmetrization} $\widetilde{T}$ as
the $(r,k)$-tensor derived from $T$ by symmetrizing its $k$ contravariant legs:
$$\widetilde{T}(v_1, \dots, v_k)=\frac1{k!}\sum_{\sigma\in S_k}T(v_{\sigma(1)}, \dots, v_{\sigma{k}}). $$
If we need to symmetrize only a subset of the 
contravariant legs, this will be indicated in some explicit fashion. 
For example, if $T$ is an $(r, k+l)$-tensor,
we separate by a semicolon the vector  arguments that will not be symmetrized
and place them last in order of insertion:
$ T(v_1, \dots, v_k; w_1, \dots, w_l).$
 Diagrammatically, the 
input legs taking the arguments $w_1, \dots, w_l$
could be shown, for example,  on either side of the tensor box,
although we will simply omit them in our diagrams. 
A
thick line crossing the other contravariant legs is added to indicate 
symmetrization.  
 
 \vspace{.1in}

It is convenient to introduce a binary operation on  
tensors, which we call {\em symmetric composition}.
Let $Q$ be an $(r, s)$-tensor $Q$ and $H$ a $(l, k)$-tensor, where $l\leq s$. 
We define the symmetric $(r, s+k-l)$-tensor $Q\odot H$ as follows:
$$ Q\odot H(v_1, \dots, v_{s+k-l})=\frac1{(s+k-l)!}\sum_{\sigma\in S_{s+k-l}}Q(H(v_{\sigma(1)}, \dots, v_{\sigma(k)}), v_{\sigma(k+1)}, \dots, v_{\sigma(s+k-l)})$$
We have assumed  here for simplicity that all the contravariant legs are 
involved in the symmetrization. The more general case  is defined similarly.  

 \vspace{.1in}
\begin{figure}[htbp]
\begin{center}
\includegraphics[width=1in]{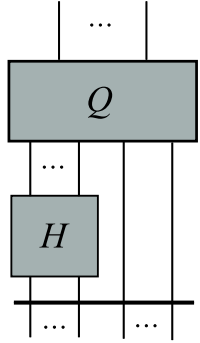}
\caption{\small  Symmetric composition $Q\odot H$ of $Q$ and $H$. }
\label{sym} 
\end{center}
\end{figure}

It is clear from the diagram that $Q\odot H$ would still make sense if $l> s$, in which case
the result of the operation is a symmetric $(r+ l-s, k)$-tensor. Below we mainly need the
case  where $l=1$.  We also note that 
 the input legs to which symmetrization is typically applied
 later in the paper  
are created   from   multiple applications of  covariant differentiation to a tensor.

\begin{propo}
The symmetric composition of $Q$ and $H$ satisfies the product rule:
$$ {D} _v (Q\odot H)= ({D}_v Q)\odot H + Q\odot {D}_v H.$$
\end{propo}
\begin{pf}
The proof is a tedious but straightforward application of Equation \ref{tensoreq} and
the easily proved fact that 
the $(k,k)$-tensor $P:v_1\otimes \dots \otimes v_k\rightarrow v_{\sigma(1)}\otimes \dots \otimes v_{\sigma(k)}$ satisfies ${D} P=0$,   where $\sigma$ is any  permutation. 
\end{pf}

When $Q$ and $H$ are themselves symmetric, $Q\odot H$  has the following useful description.
Let $\mathcal{C}(k,l)$ represent the collection of all $l$-subsets of $\{1, \dots, k\}$ (i.e., subsets
with $l$ elements).  The cardinality of $\mathcal{C}(k, l)$ is the binomial coefficient
$\binom{k}{l}=k!/(k-l)!l!$.  Let $\pi:S_k\rightarrow \mathcal{C}(k,l)$ denote the map that
associates to each permutation $\sigma$ the set $J=\{\sigma(1), \dots, \sigma(l)\}$, and
write $S_J=\pi^{-1}(J)$. Then each $S_J$, $J\in \mathcal{C}(k, l)$, has cardinality $l!(k-l)!$
and $S_k$ is partitioned into the disjoint union of the $S_J$. 
The following notation will be used: let $v_1, \dots, v_k$
be a subset of $T_x\mathbb{R}^n$, $H$  a symmetric $(m,l)$-tensor at $x$, and  $J$
an $l$-subset of $\{1, \dots, k\}$. Then it makes sense to write
$ H(J):=H(v_{j_1}, \dots, v_{j_l})$ for $J=\{j_1, \dots, j_l\}.$
If $Q$ is a symmetric $(r,s)$-tensor, where $s\geq m$ and $s-m=k-l$,
we similarly write
$Q(H(J), J^c) $, where $J^c$ indicates the complement of $J$.
(More generally, $Q$ and $H$ may contain additional contravariant legs,
with respect to which the tensors are not necessarily symmetric, 
that are not included in the operation of  symmetric composition.)

 \vspace{.1in}
\begin{figure}[htbp]
\begin{center}
\includegraphics[width=5in]{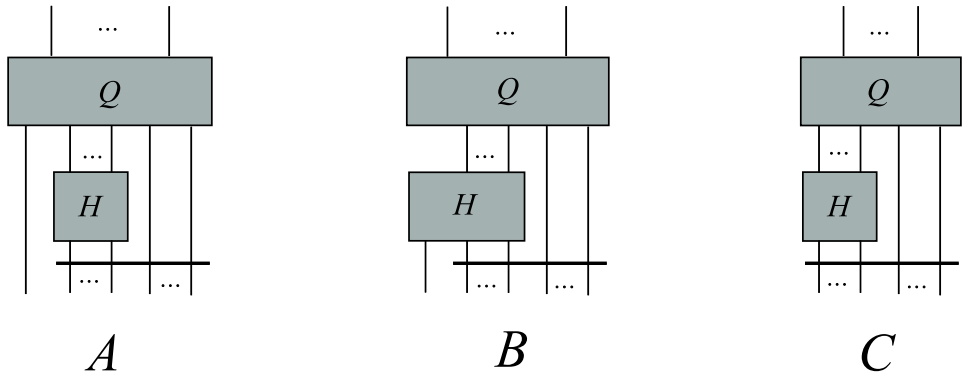}
\caption{\small  
Tensors $A, B, C$ satisfying the 
  identity 
$\binom{k-1}{l+1}A+\binom{k-1}{l}B= \binom{k}{l+1}C$ of
 Proposition \ref{identity_propo}.}
\label{identity} 
\end{center}
\end{figure}

If $v\in T_x\mathbb{R}^n$ and $H$ is an $(l, k)$-tensor,   let
 $i_vH$ denote  the $(l, k-1)$-tensor such that $(i_vH)(v_2, \dots, v_k)=H(v, v_2, \dots, v_k)$.
 \begin{propo}\label{identity_propo}
 Let $H$ be an $(u,l+1)$-tensor, $Q$ an $(m, k-l-1+u)$-tensor, both symmetric
 in their lower legs, and $v\in T_x\mathbb{R}^n$.
 Then 
 $$\binom{k}{l+1} i_v(Q\odot H)=\binom{k-1}{l+1} i_vQ\odot H + \binom{k-1}{l}Q\odot i_v H.$$
 \end{propo}
\begin{pf}
Let $v_1=v$, $v_2, \dots, v_k$ be elements of $T_x\mathbb{R}^n$. The key point
 to observe is that
 the sum of  $Q(H(J), J^c)$ over all $J\in \mathcal{C}(k, l+1)$ is equal
 to the sum over all $J\in \mathcal{C}(k,l+1)$ that   contain $1$ plus
 the sum over all $J$ that do not.
\end{pf}

\subsection{Identities for the Lie derivative}
The infinitesimal action of the flow $\phi_t$ on tensors is given by the Lie derivative with
respect to the vector field $F$. We register here for later use some useful formulas  involving the
Lie and covariant derivatives.
We  recall that  the Lie  derivative of tensor fields is defined just
as we did for the covariant derivative, except that, on a vector field $X$, it is given
by the Lie bracket $\mathcal{L}_FX=[F,X]$, and on functions $\mathcal{L}_Ff=Ff$. Another way to describe $\mathcal{L}_FT$ is as the time
derivative of the tensor defined by ``convecting'' $T$ along the flow of
$F$. 

Let now $F$ be a  smooth vector field   (thus $F$ is
a $(1,0)$-tensor) and  set  $F^{(k)}={D}^kF$. Thus $F^{(k)}$ is a $(1,k)$-tensor.
For the moment, we only make use of $F^{(1)}$. 

The covariant and Lie derivatives of tensors with respect to a vector field $F$ are
related via an ``algebraic derivative'' with respect to $F^{(1)}$. The latter operation
is defined as follows. Let $B$ be a field of linear maps and define $\mathcal{A}_B$
by the properties: (1) $\mathcal{A}_Bf=0$ on functions; (2) $\mathcal{A}_Bv=Bv$
on vectors; (3) $\mathcal{A}_B\alpha=-\alpha \circ B$ on covectors;  (4) $\mathcal{A}_B$
satisfies the product rule with respect to tensor product.  It can be checked that
these properties uniquely determine a linear map on tensors. In particular,
  if $Q$ is a $(0,k)$-tensor, 
$$ (\mathcal{A}_BQ)(v_1, \dots, v_k)=-\sum_{j=1}^k Q(v_1, \dots, Bv_j, \dots, v_k).$$
From   the properties stated in  Proposition \ref{lieeq} below we obtain
that 
 if $Q$ is a  $(0, k)$-tensor field, then
for any point $x$ and tangent  vectors $v_1, \dots, v_k$ at $x$,
$$(\mathcal{L}_FQ)_x(v_1, \dots, v_k)=({D}_FQ)_x(v_1, \dots, v_k) + \sum_{j=1}^k Q_x(v_1, \dots, F^{(1)}(v_j), \dots, v_k), $$
 and if moreover $Q$ is symmetric, then
$ \mathcal{L}_FQ={D}_FQ  + k Q\odot F^{(1)}.$
The next proposition summarizes some relations among the various derivative operations.
Here $[\cdot, \cdot]$ is the standard commutator of operators. 
Also note that, 
if we  regard $F$ as generating the time 
evolution of a system, then   with
  a slight abuse of notation,  
${D}_FQ=\dot{Q}$, where the dot over $Q$ means time derivative.
Below, $F^{(l+1)}(X_1, \dots, X_l, \cdot)$ is the field of linear maps defined by inserting
a vector into  the last slot.

\begin{propo}\label{lieeq}
Let  $F$, $X$, $X_1, \dots, X_l$   be  vector fields and $Q$ an $(r,k)$-tensor field. Then
\begin{enumerate}
\item 
${D}_FQ-\mathcal{L}_FQ = \mathcal{A}_{F^{(1)}}Q$
\item   $[{D}_X, \mathcal{A}_{F^{(1)}}]Q= \mathcal{A}_{F^{(2)}(X, \cdot)}Q$
\item $[{D}_{X_1}, \mathcal{A}_{F^{(l)}(X_2, \dots, X_{l}, \cdot)}]=\mathcal{A}_{F^{(l+1)}(X_1, \dots, X_{l}, \cdot) +\sum_{j=2}^{l-1}F^{(l)}(X_2, \dots, {D}_{X_1}X_j, \dots, X_{l}, \cdot)}$
\item $[\mathcal{L}_F, {D}_X]Q={D}_{[F,X]}Q+ \mathcal{A}_{F^{(2)}(X, \cdot)}Q$
\item $(\mathcal{L}_F Q^{(1)})(X;\cdot)={D}_X\mathcal{L}_FQ+\mathcal{A}_{F^{(2)}(X, \cdot)}Q.$
\end{enumerate}
\end{propo}
\begin{pf}
It is easily checked that ${D}_F-\mathcal{L}_F$   has the  properties
defining $\mathcal{A}_{F^{(1)}}$, so   1 is a consequence of uniqueness of
$\mathcal{A}_{F^{(1)}}$. A similar verification also proves 2. Property 3 is shown
by induction and the same argument used for 1 and 2 based on uniqueness of the
algebraic derivative. 
Properties 4 and 5 can be proved using 1 and  2 by a tedious but straightforward
algebraic manipulation. When deriving these properties, it should be born in mind 
that ${D}$ is torsion-free and flat. 
\end{pf}

 Property 5 of the above proposition gives a way
of finding $\mathcal{L}_FQ^{(k)}$ recursively 
if $\mathcal{L}_FQ$ is  known.
We illustrate this  with 
 a  formula for   $\mathcal{L}_FQ$  when
  $Q=f^{(k)}$ and $f$ is
a function. This is the case we need to consider 
in extending Malkin's method. The key point  to notice   is that
the formula  expresses $\mathcal{L}_Ff^{(k)}$ in terms of the lower order
tensors $f^{(l)}$, $l=1, \dots, k-1$.

\begin{propo}\label{formula}
Let $F$ be a smooth  vector field and $f$ a smooth function.
Suppose that $Ff=g$, for some smooth function $g$.
 Then $\mathcal{L}_F f^{(1)}=g^{(1)}$ and, for $k\geq 2$,
$$ \mathcal{L}_Ff^{(k)}=g^{(k)}-\sum_{l=2}^k \binom{k}{l} f^{(k+1-l)}\odot F^{(l)}.$$ 
\end{propo}
\begin{pf}
The proof is by induction. Using Property 5 of Proposition \ref{lieeq}
one immediately gets $\mathcal{L}_Ff^{(1)}=g^{(1)}$ and
\begin{align*}
\left(\mathcal{L}_Ff^{(2)}\right)(v_1,v_2)&=\left({D}_{v_1}\mathcal{L}_F f^{(1)}\right)(v_2) +\left(\mathcal{A}_{F^{(2)}(v_1, \cdot)}f^{(1)}\right)(v_2)\\
&=\left({D}_{v_1}g^{(1)}\right)(v_2)-f^{(1)}\left(F^{(2)}(v_1, v_2)\right)\\
&=\left(g^{(2)} - f^{(1)}\odot F^{(2)}\right)(v_1, v_2).
\end{align*}
So we suppose that 
the equation holds for $k\geq 2$ and wish to obtain it for $k+1$. 
First note that
\begin{equation*}
\begin{split}
\left({D}_{v_1}\mathcal{L}_F f^{(k)}\right)(v_2, \dots, v_{k+1})&={D}_{v_1}\left(g^{(k)}-\sum_{l=2}^k\binom{k}{l}f^{(k+1-l)}\odot F^{(l)}\right)(v_2, \dots, v_{k+1})\\
&=g^{(k+1)}(v_1, \dots, v_{k+1})\\
&\qquad
-\sum_{l=2}^k\binom{k}{l}\left({D}_{v_1}f^{(k+1-l)}\odot F^{(l)}\right)(v_2, \dots, v_{k+1})\\
&\qquad\qquad -\sum_{l=2}^k\binom{k}{l}\left(f^{(k+1-l)}\odot {D}_{v_1}F^{(l)}\right)(v_2, \dots, v_{k+1})\\
&=g^{(k+1)}(v_1, \dots, v_{k+1})\\
&\qquad
-\sum_{l=1}^{k-1}\binom{k}{l+1}\left({D}_{v_1}f^{(k-l)}\odot F^{(l+1)}\right)(v_2, \dots, v_{k+1})\\
&\qquad\qquad -\sum_{l=1}^{k-1}\binom{k}{l}\left(f^{(k+1-l)}\odot {D}_{v_1}F^{(l)}\right)(v_2, \dots, v_{k+1})\\
&\qquad\qquad\qquad -\left(f^{(1)}\odot F^{(k+1)}\right)(v_1, \dots, v_{k+1})\\
&\qquad\qquad\qquad\qquad +k\left(f^{(k)}\odot {D}_{v_1}F^{(1)}\right)(v_2, \dots, v_{k+1}).
\end{split}
\end{equation*}
Using the tensor identity of Proposition \ref{identity_propo},
the above simplifies to
\begin{equation*}
\begin{split}
\left({D}_{v_1}\mathcal{L}_F f^{(k)}\right)(v_2, \dots, v_{k+1})&=g^{(k+1)}(v_1, \dots, v_{k+1})\\
&\qquad
-\sum_{l=2}^{k+1}\binom{k+1}{l}\left(f^{(k+2-l)}\odot F^{(l)}\right)(v_1, \dots, v_{k+1})\\
&\qquad\qquad +k\left(f^{(k)}\odot {D}_{v_1}F^{(1)}\right)(v_2, \dots, v_{k+1}).
\end{split}
\end{equation*}

By Property 5 of Proposition \ref{lieeq}, 
\begin{equation*}
\begin{split}
\mathcal{L}_Ff^{(k+1)}(v_1, \dots, v_{k+1})&=\left({D}_{v_1}\mathcal{L}_F f^{(k)}\right)(v_2, \dots, v_{k+1}) + \left(\mathcal{A}_{F^{(2)}(v_1, \cdot)}f^{(k)}\right)(v_2, \dots, v_{k+1})\\
&=\left({D}_{v_1}\mathcal{L}_F f^{(k)}\right)(v_2, \dots, v_{k+1})  \\
&\qquad  -\sum_{j=2}^{k+1} f^{(k)}(v_2, \dots, F^{(2)}(v_1, v_j), \dots, v_{k+1})\\
&=g^{(k+1)}(v_1, \dots, v_{k+1})\\
&\qquad
-\sum_{l=2}^{k+1}\binom{k+1}{l}\left(f^{(k+2-l)}\odot F^{(l)}\right)(v_1, \dots, v_{k+1})\\
&\qquad\qquad +k\left(f^{(k)}\odot {D}_{v_1}F^{(1)}\right)(v_2, \dots, v_{k+1})\\
&\qquad\qquad\qquad -\sum_{j=2}^{k+1}f^{(k)}(v_2, \dots, F^{(2)}(v_1,v_j), \dots, v_{k+1})\\
&=g^{(k+1)}(v_1, \dots, v_{k+1})\\
&\qquad
-\sum_{l=2}^{k+1}\binom{k+1}{l}\left(f^{(k+2-l)}\odot F^{(l)}\right)(v_1, \dots, v_{k+1}),
\end{split}
\end{equation*}
which is the claimed formula for $k+1$.
\end{pf}

The case of main interest  is $f=\Theta$, for which $F\Theta=1/T$ is constant.
Thus $g^{(k)}=0$ for $k\geq 1$ and we obtain
\begin{equation}\label{main}
\mathcal{L}_F\Theta^{k}=-\sum_{l=2}^{k}\binom{k}{l}\Theta^{(k+1-l)}\odot F^{(l)}.
\end{equation}

\subsection{Proof of Theorem \ref{main_theorem}}
We turn now to the proof of Theorem \ref{main_theorem}.
Part (1) of the theorem, which gives the differential equation satisfied by $\Theta^{(k)}$ 
is restated  in the next proposition.
 Recall the notation:
$ Q^{(k)}_t = \Theta^{(k)}_{\phi_t(x)},$
where $\phi_t$ is the flow of $F$ and $x$ is arbitrary. Indicating the time derivative of $Q^{(k)}$ 
by $\dot{Q}^{(k)}$, we have
$\dot{Q}^{(k)}_t=({D}_F\Theta^{(k)})_{\phi_t(x)}.$

\begin{propo}\label{dfq}
We assume the notation of Section \ref{main_result}. Let $Q^{(k)}$ be as just defined. Then
$$\dot{Q}_t^{(k)} + k Q_t^{(k)}\odot F^{(1)}= - \sum_{l=1}^{k-1} \binom{k}{l+1}Q_t^{(k-l)}\odot F^{(l+1)}.$$
\end{propo}
\begin{pf}
According to Proposition \ref{lieeq}, ${D}_F\Theta^{(k)} -\mathcal{A}_{F^{(1)}}\Theta^{(k)}=\mathcal{L}_F\Theta^{(k)}$. On the other hand,  $\mathcal{A}_{F^{(1)}}\Theta^{(k)}=-k\Theta^{(k)}\odot F^{(1)}$, and
since the directional derivative  $F\Theta$ is constant, 
$\mathcal{L}_F\Theta^{(k)}$ is given by Equation \ref{main}.  
We get the claimed formula by finally rewriting the
resulting equation in terms of $Q^{(k)}_t$.
\end{pf}

 In what follows, 
 let $W_t$ be the tangent space to the isochron at $\phi_t(x)$, $x\in \mathcal{C}$.
 Vectors in $W_t$, by assumption,  contract exponentially under the flow; i.e.,  
  $|d\phi_s v| < C\lambda^s|v|$ for all $v\in W_t$ and  positive constants $C$ and
  $\lambda<1$, where $s\geq 0$.  If $v$ is parallel to $F(\phi_t(x))$ then
  $|d\phi_s v|$ is bounded above as well as away from $0$.  We similarly need to know the decay  properties of
  tensors of type $(0,k)$ under the flow. 
  The natural push-forward action of $\phi_s$
  on 
 a tensor-valued
  function, $\tau_t$,    of type $(0,k)$   along $\mathcal{C}$
  is defined by
  $$(\phi_s\cdot \tau_t)(u_1, \dots, u_k)=\tau_t((d\phi_s)_y^{-1}u_1, \dots,
  (d\phi_s)_y^{-1}u_k),$$
  where $u_1, \dots, u_k$ are   vectors  at $y$. 
  This applies, in particular, to a tensor field $\tau$ defined in a neighborhood
  of $\mathcal{C}$, in which case $\tau_t=\tau_{\phi_t(x)}$ is a $T$-periodic tensor-valued
  function of $t$. The function is flow-invariant if $\phi_s\cdot \tau_t=\tau_{t+s}$.

    Let $\mathbb{R}\tau$ represent the one-dimensional space over $\mathbb{R}$
    spanned by a tensor $\tau$. Since  the family of  isochrons   and the vector
    field $F$ are invariant under the flow,
   the tangent space to $\mathbb{R}^n$ decomposes
  invariantly as a direct sum
  $$T_{\phi_t(x)}\mathbb{R}^n=\mathbb{R} F_t\oplus W_t,$$ 
  where $F_t=F(\phi_t(x))$. Let $W_t^*$
  be the subspace of $T^*_{\phi_t(x)}\mathbb{R}^n$ consisting of covectors
that vanish on $F_t$. It is not difficult to check that
$$T^*_{\phi_t(x)}\mathbb{R}^n=\mathbb{R}Q^{(1)}_t\oplus W_t^*$$
  is also a flow-invariant decomposition. By general tensor algebra, one
  also obtains  a flow-invariant decomposition  of the space 
  of $(0,k)$-tensors as a direct sum  of subspaces of the form
  $$V_{k_1, k_2, t}=\mathbb{R} Q_t^{(1)}\otimes \cdots \otimes \mathbb{R} Q_t^{(1)}\otimes
  W_t^*\otimes \cdots \otimes W_t^*,$$
  in which there are $k_1$ copies of $\mathbb{R} Q_t^{(1)}$ and $k_2$ copies
  of $W_t^*$, where $k_1+k_2=k$ and $0\leq k_1\leq  k$.

 The Euclidean norm on vectors, $|v|$, extends  in
    natural ways to norms on tensors of any kind. We use the same notation,
    $|\tau|$, for the norm of a tensor $\tau$ of general type.
    We refer the reader to
    texts on multilinear algebra  or differential geometry for how this    can 
    be defined, although it is not necessary for
    what we do below to have any  explicit 
    description  in mind, and 
      the form of the theorem does not depend on a particular
        choice of norm.
      The main property     we use  below is that
    if $A$ is a $(0,k)$-tensor and $u_1, \dots, u_k$ are vectors  then
    $|A(u_1, \dots, u_k)|\leq |A|\prod_{j=1}^k |u_j|$.
 
 \begin{lem}\label{lll}
 Let    $A$ be a tensor in $V_{k_1, k_2, t}$. Then there exists a constant $C_A$  such that
 $|\phi_s\cdot A|\leq C_A \lambda^{k_2|s|}$
 for all $s<0$. 
 \end{lem}
 \begin{pf}
 A tensor in $V_{k_1, k_2, t}$ has the form $ Q_t^{(1)}\otimes \cdots \otimes Q_t^{(1)}\otimes \widetilde{A}$ where $\widetilde{A}$ belongs to the $k_2$-fold tensor power of
 $W_t^*$. Thus $|\phi_s\cdot A|\leq K |\phi_s\cdot \widetilde{A}|$,
 where $K$ is the $k_1$th power  of the supremum of $|Q_s^{(1)}|$ over the period $0\leq s\leq T$.
Since $ \widetilde{A}$ vanishes whenever any one of its arguments is parallel to $F_{t}$,
 the norm of $\phi_s\cdot \widetilde{A}$  can be bounded
 above by the supremum of 
 $|\widetilde{A}((d\phi_s)^{-1}u_1, \dots, (d\phi_s)^{-1}u_1)|$
 over all $u_j\in W_{t+s}$ of norm at most  $1$. 
 Now, 
 $$\left|\widetilde{A}((d\phi_s)^{-1}u_1, \dots, (d\phi_s)^{-1}u_1)\right|\leq \left|\widetilde{A}\right|\prod_{j=1}^{k_2}
 \left|(d\phi_s)^{-1}u_j\right|\leq C^{k_2}\lambda^{k_2|s|}$$
 which proves the assertion of the lemma. 
 \end{pf}

 \begin{propo} \label{xx}
 Let $x\in \mathcal{C}$ and  $A_t^{(k)}$
  a $(0, k)$-tensor at $\phi_t(x)$  depending differentiably
 on $t$.
   Suppose that $A^{(k)}_t$ is a   solution   of the homogeneous equation 
   $$\dot{A}_t^{(k)} + k A^{(k)}_t\odot F^{(1)}=0.$$ Then 
    for some positive constant $K$ and for $\lambda < 1$ as above, there exists $c$ such that 
    $$\left|A_t^{(k)}-c \ \! Q_t^{(1)}\otimes \cdots \otimes Q_t^{(1)}\right|<K\lambda^{|t|}$$ 
    for all  $t<0$, where the tensor product  on the left-hand side  contains $k$ terms. In particular, if $A_t^{(k)}$ is   $T$-periodic, then
    $A_t^{(k)}=c \ \! Q_t^{(1)}\otimes \cdots \otimes Q_t^{(1)}$.
 \end{propo}
 \begin{pf}
 Since $F_x=\frac{d}{dt}$ along the limit cycle  (by the usual identification of a vector field as a derivation), 
   $\dot{A}_t$ equals ${D}_F A$, where the latter has to be
  interpreted in terms  of an arbitrary extension of $A$ on a neighborhood
  of the limit cycle. Now, according to part (1) of  Proposition \ref{lieeq}, 
  $$\mathcal{L}_F A_t^{(k)}=\dot{A}_t^{(k)} -\mathcal{A}_{F^{(1)}}A_t^{(k)}=\dot{A}^{(k)} + k A^{(k)}\odot F^{(1)}=0$$
  on $\mathcal{C}$.
  Therefore,  the proposition amounts to the assertion that (1) the space of periodic, flow-invariant
  tensor-valued functions of type $(0,k)$ on $\mathcal{C}$ is one-dimensional,
  spanned by the $k$th tensor power of  $Q_t^{(1)}$, and (2) if $A_t^{(k)}$ is
  flow-invariant but not necessarily periodic, the stated inequality holds, i.e., 
  the components of a  solution   transverse
  to the one-dimensional space of  periodic solutions must contract exponentially 
  to $0$.
   
   Thus suppose first that $A_t^{(k)}$ is flow-invariant and $T$-periodic. 
   Let $A_t^{(k_1, k_2)}$ be the component of $A_t^{(k)}$ in $V_{k_1, k_2, t}$.
   In particular, $A^{(k,0)}_t = c_t Q_t^{(1)}\otimes \cdots \otimes Q_t^{(1)}$.
   Since the subspaces $V_{k_1, k_2, t}$ are flow-invariant and periodic, each component
   $A_t^{(k_1, k_2)}$ is also flow-invariant and periodic.
    In particular, $c_t$ is constant (independent of $t$) and
    $A_t^{(k_1, k_2)}=0$ whenever $k_2\neq 0$, by Lemma \ref{lll}.
    This shows claim (1). To verify claim (2) we suppose that 
    $A_t^{(k)}$ is flow-invariant but not necessarily periodic.  Thus
    $A_{t+s}^{(k)} = \phi_s\cdot A_t^{(k)}$. Now each component in the
    decomposition
    $$A_t^{(k)}=c_t Q_t^{(1)}\otimes \cdots \otimes Q_t^{(1)} + \sum_{k_2\neq 0} A_t^{(k_1,k_2)} $$
    is flow-invariant, so $c_{t+s}=c_t$ for all $s$ and 
    \begin{align*}\left|A_{t+s}^{(k)}-c_t Q_{t+s}^{(1)}\otimes \cdots \otimes Q_{t+s}^{(1)} \right| &\leq
     \sum_{k_2\neq 0} \left|A_{t+s}^{(k_1, k_2)}\right|\\ 
    &= \sum_{k_2\neq 0} \left|\phi_s\cdot A_{t}^{(k_1, k_2)}\right|\\ 
    &\leq K \lambda^{|s|}
    \end{align*}
    for some constant $K$, where we have used again Lemma \ref{lll}. This concludes the
    proof of the proposition.
 \end{pf}

 We can now prove parts (2) and (3) of Theorem \ref{main_theorem}.
For $k=1$, the equation reduces to  $$\dot{Q}_t^{(1)} + Q_t^{(1)}\circ F^{(1)}=0.$$
This equation can be solved numerically by the standard (first order) approach,
which essentially amounts to Proposition  \ref{xx} in the special
case $k=1$: Fix $x\in \mathcal{C}$ and
let $A$ be a choice
of  
initial value,  arbitrary except for  the condition $A(F_x)=1/T$.
One then integrates the first order equation for $t<0$ for large enough values of $|t|$ until
the solution stabilizes to a periodic matrix-valued function over the limit cycle. 
That function is the sought after solution for  $Q_t^{(1)}$. For larger values of $k$,
we regard the differential equation in Proposition \ref{dfq} as non-homogeneous:
$$\dot{Q}_t^{(k)} + k Q^{(k)}_t\odot F^{(1)}= B^{(k)}_t,$$
where the right-hand side is assumed to have already been obtained in
the previous steps.

We proceed by induction.  Suppose
that we have found $Q^{(1)}_t, \dots, Q^{(k-1)}_t$ and wish to obtain
$Q^{(k)}_t$. 
Set \begin{equation}\label{gen} B_t^{(k)}= - \sum_{l=1}^{k-1} \binom{k}{l+1}Q_t^{(k-l)}\odot F^{(l+1)}
\end{equation}
and integrate the equation 
$$\dot{A}_t^{(k)} + k A^{(k)}_t\odot F^{(1)}= B^{(k)}_t$$
for $t\leq 0$,
starting with an arbitrary initial condition,
 until  ${A}_t^{(k)}$ stabilizes 
to a periodic function of $t$. 
Stabilization must occur at the exponential rate given by Proposition \ref{xx},
since the difference ${A}_t^{(k)}-Q_t^{(k)}$
is a solution of the homogeneous equation of Proposition \ref{xx}.
This  periodic solution is still denoted
by $A_t^{(k)}$ (in particular, $A^{(k)}_0$ is  the
value at $x$ after stabilization). The true solution we seek, $Q^{(k)}_t$, differs from
$A_t^{(k)}$   by a periodic homogeneous solution so there exists
a constant $c$ such that 
$$ Q^{(k)}_t= A^{(k)}_t + c\ \! Q_t^{(1)}\otimes \cdots \otimes Q_t^{(1)}.$$
Since $Q^{(1)}_t(F)=1/T$, the constant $c$ has the form
$$ c=T^{k}[ Q_0^{(k)}(F_x, \dots, F_x)-A^{(k)}_0(F_x, \dots, F_x)].$$

It remains to argue that $Q_0^{(k)}(F_x, \dots, F_x)$ can be expressed
in terms of the lower order $Q^{(l)}_0$, $F$, and its derivatives at $x$.
Recall that $Q_t^{(k)}=\Theta^{(k)}_{\phi_t(x)}$. The following general fact holds for
$\Theta^{(k)}_y$ (for $y$ not necessarily on the limit cycle). We set  $F^{(0)}=F$.
\begin{propo} \label{forestt}
 There exists for each $k$ an algebraic  function 
that gives  $\Theta^{(k)}_y(F_y, \dots, F_y)$
in terms of the 
$\Theta^{(l)}_y$, for $l=1, \dots, k-1$, and the $F^{(j)}_y$, for $j=0, \dots, k-1$,
 and  $y$
  in   a neighborhood of $\mathcal{C}$.  
\end{propo}

The precise meaning of this proposition, and an algorithm for obtaining the
indicated algebraic function, are explained and illustrated in the next section.

\subsection{Forest expansion of $F^k\Theta$}\label{forest}
We wish to expand $\Theta^{(k)}(F, \dots, F)$, at any given point,  in
terms of the $\Theta^{(j)}$ and $F^{(j)}$  for $j<k$, 
and thus find the algebraic function indicated in Proposition \ref{forestt}.
This can   naturally be done by successive applications of
the chain rule, applied to the expression $F^k\Theta=0$, $k\geq 2$.
(Recall that $F\Theta=\Theta^{(1)}(F)=1/T$.) What is needed is 
a method    to conveniently deal with the
combinatorial task of enumerating the terms of this expansion.
 This can be done by enumerating rooted trees with $k$ edges.
We describe here this {\em forest expansion} method and 
illustrate it  with a few examples. 

Finding the general form of the algebraic function claimed in 
Proposition \ref{forest} for an
arbitrary $k$ is  a complicated combinatorial problem, and amounts to
a type of Fa\`a di Bruno formula,  which we do not attempt to describe here.
We are content with giving the forest expansin algorithm  and
 applying it to small values of $k$.

\begin{figure}[htbp]
\begin{center}
\includegraphics[width=.8in]{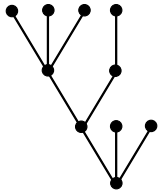}
\caption{\small  A rooted tree diagram representing  
$ \Theta^{(3)}(F^{(2)}(F^{(3)}(F, F, F), F^{(1)}(F)), F, F)$,
which is one term in the expression of $F^9\Theta$. The tree itself will be represented by
$\tau_3(\tau_2(\tau_3,\tau_1), \tau_0,\tau_0)$. }
\label{tree} 
\end{center}
\end{figure}

Figure \ref{tree} explains  how to represent 
  nested evaluations of tensors
by rooted trees.
 The number of edges of the rooted tree diagram
is the total order of differentiation, so all diagrams associated to $F^k\Theta$ will
contain $k$ edges. The degree of the root vertex is the order of differentiation
of $\Theta$ (this is $3$ in the example of Figure \ref{tree}), and all the other vertices represent a $F^{(j)}$,
where $j+1$ is the vertex degree. Thus to each   leaf (i.e.,  a
vertex  with no descendants) is attached an $F$, and to 
 each non-root vertex is recursively attached a vector
as follows: Starting from the   leaves (associated to copies of $F$)
one moves down one step to the parent vertices (associated to copies of $F^{(j)}$,
if $j+1$ is the degree of a parent vertex) and compute  
$F^{(j)}(F, \dots, F)$. This vector is   now attached to each of those second-to-last   generation vertices.
These new vectors in turn are  evaluated into the tensors attached
to their parent vertices. 
We continue this process until the vectors attached to the first generation vertices (the
ones connected to the root by an edge) are evaluated into
 $\Theta^{(l)}$,
where $l$ is the degree of the  root vertex.

 \vspace{.1in}

The expression $\Theta^{(k)}(F, \dots, F)$ itself corresponds to a tree having
a root of order $k$ and $k$ leaves attached to it. 
We denote this tree by $\tau_k$,
$k\geq 0$, where $\tau_0$ consists of a vertex with $0$ edges, i.e., a leaf.
Other trees are obtained by nesting trees of type $\tau_j$.
Thus each $\tau_j$ can take $j$ arguments, each of which is a  tree  of the same kind
(for possibly different $j$). For
example, 
$\tau_l(\tau_{j_1}, \dots, \tau_{j_l})$ represents  a tree that consists of a 
root vertex of degree $l$ and at the non-root vertex of each of the $l$ edges is appended 
the  tree $\tau_{j_s}$ so that the root vertex of the latter is identified with the
 non-root vertex of the former.

It is clear that $F^k\Theta$ can in general be represented by
a forest of rooted trees with $k$ edges, each tree being assigned  some multiplicity. We will
see shortly how the multiplicities are 
determined by  
counting the ways a tree is derived from other 
trees with $k-1$ edges.
Figure \ref{forest} shows the forest diagram representation of
$F^2\Theta$, $F^3\Theta$ and $F^4\Theta$.

\begin{figure}[htbp]
\begin{center}
\includegraphics[width=4.5in]{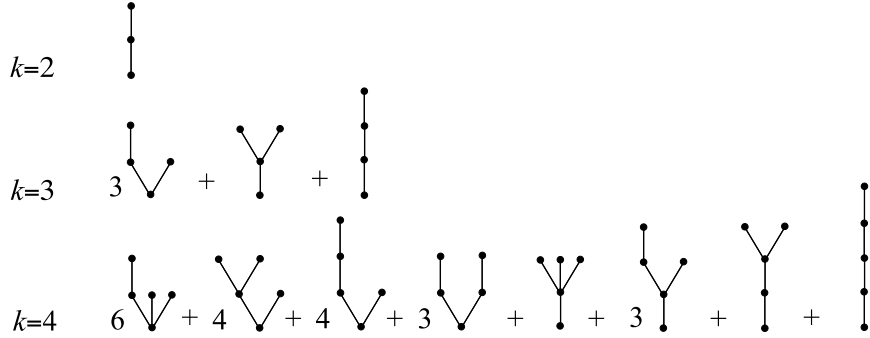}
\caption{\small  The forest representation of $-\Theta^{(k)}(F, \cdots, F)$ for $k=2,3,4$. }
\label{forest4} 
\end{center}
\end{figure}

We give a few  examples of the forest expansion before showing the general method.  
By the basic rules of covariant differentiation of
tensors, we obtain
$FF\Theta= F\Theta^{(1)}(F)=\Theta^{(2)}(F, F)+ \Theta^{(1)}(F^{(1)}(F)). $
Therefore,
$$ \Theta^{(2)}(F, F)=- \Theta^{(1)}(F^{(1)}(F)).$$
For $k=3$, we have:
\begin{align*} FFF\Theta&= F[\Theta^{(2)}(F, F)+ \Theta^{(1)}(F^{(1)}(F))]\\
&=\Theta^{(3)}(F,F,F) + 3\Theta^{(2)}(F^{(1)}(F),F)+\Theta^{(1)}(F^{(2)}(F,F))+
\Theta^{(1)}(F^{(1)}(F^{(1)}(F))),
\end{align*}
so that 
\begin{equation*}
-\Theta^{(3)}(F,F,F)= 3\Theta^{(2)}(F^{(1)}(F),F)+\Theta^{(1)}(F^{(2)}(F,F))+
\Theta^{(1)}(F^{(1)}(F^{(1)}(F))).
\end{equation*}
The expansion of $-\Theta^{(4)}(F,F,F,F)$ is shown diagrammatically
in Figure \ref{forest4}.

 \vspace{.1in}

It should now be apparent  that the algebraic expression giving 
$-\Theta^{(k)}(F, \dots, F)$ as a function of the lower order terms $\Theta^{(j)}$
and the $F^{(l)}$, as claimed in Proposition \ref{forest},
is precisely the forest expansion of $-\Theta^{(k)}(F, \dots, F)$.
It is also  clear that all  rooted trees with $k$ edges (except $\tau_k$), up to isomorphism,  
appear as a term  in the forest expansion  of $\tau_k$, for a given
$k$. What is needed then is 
   a more formal description of how to determine the integer  coefficients
  of  the expansion.  (Notice our slight abuse of language
  in referring   to the forest expansion of
  $F^k\Theta$ or of $-\Theta^{(k)}(F, \dots, F)$ as the  same thing. Of course, 
  the expansion of the former contains one extra term, which is (minus) the latter.)
  
   Let $\mathcal{T}(k)$ denote the set of (isomorphism classes of)
rooted trees with $k$ edges, $k\geq 0$, and denote by $m:\mathcal{T}(k)\rightarrow \mathbb{N}\cup \{0\}$
the {\em multiplicity function}, which assigns to each tree its 
coefficient  in the forest expansion of 
 $F^k\Theta$.   As  already defined, each
tree gives rise to a number: to the root vertex we associate
$\Theta^{(l)}$, where $l$ is the vertex degree; to each of the other vertices we
associate  $F^{(j)}$, where $j$ is the degree of the respective vertex and $F^{(0)}=F$;
the tensors are then evaluated as prescribed by the tree so that each vector
attached  to a vertex is an argument of the tensor attached to the parent vertex.
The result of this nested evaluation of tensors for a given $T\in \mathcal{T}(k)$ will be written $T(\Theta, F)$.
Therefore,
$$ F^k\Theta=\sum_{T\in \mathcal{T}(k)}m(T) T_p(\Theta, F).$$
 Thus, the forest expansion of $F^k \Theta$ requires an enumeration of
all rooted trees with $k$ edges, and the determination of the multiplicities $m(T)$.

A few more definitions are needed  before identifying  $m$. A tree
$T'\in \mathcal{T}(k)$ is said to {\em grow into}  
$T\in \mathcal{T}(k+1)$ if $T$ can be obtained from $T'$ by adding ({\em grafting}) one terminal edge to
any vertex of $T'$.   Conversely, 
  $T$ is {\em pruned} down  to $T'$ if $T'$ results by eliminating a terminal edge from
$T$.   Let   $\mathcal{C}_k$ be the  $\mathbb{N}$-module consisting of linear combinations over
$\mathbb{N}$ of elements of $\mathcal{T}(k)$. 
The dual of $T\in \mathcal{T}(k)$ will be written as $T^*$, so that $T^*(\sum_{S\in \mathcal{T}(k)}n_S S)=n_T.$

The {\em pruning map} $\mathcal{P}:\mathcal{T}(k+1)\rightarrow \mathcal{C}_{k}$ is defined
as follows:  For each $T\in \mathcal{T}(k+1)$ we set $\mathcal{P}(T)$ to be
the sum of all {\em distinct } trees in $\mathcal{T}(k)$ which can grow into $T$.
The {\em grafting map} $\mathcal{G}:\mathcal{C}_{k}\rightarrow \mathcal{C}_{k+1}$
is defined   on a tree $T$ by summing all trees that can be pruned down to $T$,
now counting repetitions (i.e., each tree is counted as many times
as it appears in the process of  grafting an edge at the different vertices), then multiplying the result
by $m(T)$. Now extend $\mathcal{G}$ to $\mathcal{C}_{k}$ by linearity. 
It can   be shown that the multiplicity function has the form:
$$m(T)=T^{*}(\mathcal{G}(\mathcal{P}(T))). $$
This is  based simply  on translating into diagrams
the   rule: If $A=F^{(s)}(A_1, \dots, A_s)$, then
${D}_F A=F^{(s+1)}(F, A_1, \dots, A_s) + \sum_{j=1}^s F^{(s)}(A_1, \dots, {D}_F A_j, \dots, A_s).$
The first term on the right corresponds to adding an edge to the root of the tree associated to
$A$; the other terms on the right correspond to moving up one step along one of the
root edges of $A$ and repeating the operation.

\begin{figure}[htbp]
\begin{center}
\includegraphics[width=.7in]{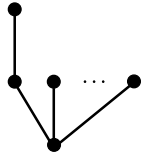}
\caption{\small  The  rooted tree $\tau_{k,1}=\tau_k(\tau_1, \tau_0, \dots, \tau_0)$. The pruning
of the single edge at distance $2$ from the root gives  $\tau_k$, 
and the pruning of any of the other edges gives $\tau_{k-1,1}=\tau_{k-1}(\tau_1, \tau_0, \dots, \tau_0)$.}
\label{example} 
\end{center}
\end{figure}

We illustrate the procedure for finding multiplicities with the example of
$\tau_k(\tau_1, \tau_0, \dots, \tau_0)$ shown in Figure \ref{example}. We use the shorter
notation $\tau_{k,1}$ for this tree.

 \vspace{.1in}

We first calculate $m(\tau_k)$. Clearly, $\mathcal{P}(\tau_k)=\tau_{k-1}$.
The grafting map gives:
\begin{align*}
\mathcal{G}(\mathcal{P}(\tau_k))&=\mathcal{G}(\tau_{k-1} )\\
&= m(\tau_{k-1})((k-1)\tau_{k-1,1}+\tau_k)\\
&= m(\tau_{k-1})\tau_k + \dots.
\end{align*}
Therefore, $m(\tau_{k})=\tau_k^*(\mathcal{G}(\mathcal{P})(\tau_k))=m(\tau_{k-1}).$
It is clear  that $m(\tau_1)=1$, so $m(\tau_k)=1$ for all $k$.  

We apply the same argument to $\tau_{k,1}$.
First, 
$\mathcal{P}(\tau_{k,1})= \tau_k +\tau_{k-1,1}. $
Now,
\begin{align*}
\mathcal{G}(\mathcal{P}(\tau_{k,1}))&=\mathcal{G}( \tau_k ) + \mathcal{G}(\tau_{k-1,1})\\
&= m(\tau_k)(k\tau_{k,1}+\tau_{k+1}) + m(\tau_{k-1,1})(\tau_{k,1} + \dots)\\
&=(km(\tau_k) +  m(\tau_{k-1,1}))\tau_{k,1} + \dots\\
&=(k+  m(\tau_{k-1,1}))\tau_{k,1} + \dots
\end{align*}
Denoting by $m_{k+1}$ the multiplicity of $\tau_{k,1}$, we obtain
$m_{k+1}= m_{k}+ k $, $m_2=1$. A simple induction gives the result:
$m(\tau_{k,1})=k(k+1)/2$.

It would be useful to derive general properties of the multiplicity map that can 
help to evaluate the forest expansion of $-\Theta^{(k)}(F, \dots, F)$. For example,
if $T$ is any rooted tree, then $$m(T)=m(\tau_1(\cdots \tau_1(T)\cdots)). $$

\end{document}